\documentclass{appolb}
\usepackage{graphicx}
\usepackage{epsfig}
\begin{document}

\title{AN OUTLOOK ON NUCLEAR PHYSICS
\thanks{Presented at the Zakopane School on Nuclear Physics "Extremes of the Nuclear Landscape", August 27 - September 2, 2012, Zakopane, Poland.} }

\author{ A.B.~Balantekin
\address{Physics Department, University of Wisconsin, Madison WI 53706 USA} \\[0.5cm]}
\maketitle

\begin{abstract}
A brief outlook on low-energy nuclear physics is presented. Selected recent developments in nuclear structure theory are highlighted and a few open questions are discussed.
\end{abstract}

\PACS{21.60.-n, 26.30.-k, 24.80.+y, 25.60.-t}
  
\section{Introduction}
Recently released decadal study report from the U.S. National Academy of Sciences, {\it Nuclear Physics: Exploring the Heart of Matter} \cite{nas1} has identified four overarching questions for nuclear physics: 
\begin{itemize}
\item How does subatomic matter organize itself and what phenomena emerge?
\item Are the fundamental interactions that are basic to the structure of matter fully understood? 
\item How did visible matter come into being and how does it evolve? 
\item How can knowledge and technological progress provided by nuclear physics best be used to benefit society?
\end{itemize}
These are challenging goals, but they represent a very special opportunity for nuclear physics. The nuclear physics community worldwide is making continuous progress towards the fulfillment of these goals. In these conference proceedings, it is impossible to cover even a fraction of this effort. I will restrict myself to highlighting selected recent developments in low-energy nuclear theory and discuss only a few of the open questions.

\section{Selected recent developments}

\subsection{Nuclear Structure theory}
Major advances in computational capabilities have pushed the boundaries of what can be calculated by microscopic {\em ab initio} 
methods. One example is the so-called Hoyle state. The 7.65 MeV $0^+$ excited state in $^{12}$C is crucial for the formation of the carbon in the Universe. Recently several authors succeeded to build this state from scratch. Epelbaum, {\em et al.} was able to carry out an 
{\em ab initio} calculation of the low-lying states of carbon-12 using supercomputer lattice simulations within the framework of effective field theory \cite{Epelbaum:2011md}. 
This state appears to have a bent-arm alpha-clustering nature. 
The Argonne group calculated the same state using Green's function Monte Carlo techniques \cite{wiringa}. 

Another example is provided by recent work on $^{14}$C. The very long lifetime of $^{14}$C as compared to other light nuclei has been an outstanding puzzle. Other light nuclei undergo beta decays with half-lives less than one day, but $^{14}$C has a beta-decay half life of $\sim$ 5,730 years, a fact that has enabled carbon-dating. Using the {\em ab initio} no-core shell model with the Hamiltonian from the chiral effective field theory, Maris {\em et al.}  found that three-nucleon force induces large cancellations  reducing the sizable contributions from the nucleon-nucleon interactions by an order of magnitude \cite{Maris:2011as}. Indeed three-nucleon interactions emerged as the salient ingredient in many recent calculations. For example, it was shown that neutron star observations can place constraints on the three-body force in neutron matter
\cite{Steiner:2011ft}. This is, in part, because the short-range three-neutron interaction determines the correlation between neutron matter energy at nuclear saturation density and higher densities relevant to neutron stars \cite{Gandolfi:2011xu}: 
Three-body forces seem to be crucial for neutron-star physics! Another recent example for the role of three-body forces was given in the context of oxygen isotopes. The neutron drip line, although evolves regularly from light to medium-mass nuclei, shows a striking anomaly in the oxygen isotopes.
Repulsive contributions to the interactions among excess neutrons due to the three-body forces resolve this anomaly \cite{Otsuka:2009cs}. 
Theoretical treatment of the three-nucleon force contributions continues to improve \cite{Hebeler:2009iv,Hebeler:2010xb}. 

Recently there have also been successful attempts to calculate nuclear reactions from first principles. One example is the {\em ab initio} no-core shell model/resonating group method approach to calculate the cross section of the $^7$Be(p,$\gamma$)$^8$B reaction. Starting from a selected similarity-transformed chiral nucleon-nucleon interaction that accurately describes two-nucleon data, Navratil {\em et al.} performed many-body calculations that simultaneously predict both the normalization and the shape of the astrophysical S-factor \cite{Navratil:2011sa}. 

Another recent development is the proper inclusion of the tensor force in the shell model Hamiltonians. For example, a new p-sd shell model Hamiltonian including up to 2-3 $\hbar \Omega$ excitations can describe the magnetic moments and Gamow-Teller (GT) transitions in p-shell nuclei well with a small quenching for the spin g-factor and the axial-vector coupling constant \cite{Suzuki:2003fw}. This new Hamiltonian significantly improves the description of the weak interactions of many nuclei for astrophysical applications.  For example it much improves the description of the cross sections for the reaction $\nu_e +^{13}$C \cite{Suzuki:2012aa}, potentially important for scintillator-based neutrino experiments. Monopole component of the nucleon-nucleon force is the same in nuclear medium and free space, however the monopole effect of the tensor force alters the shell structure in a significant way \cite{Otsuka:2005zz}. This is because the monopole component of the tensor interaction changes depending on whether the nucleon spin is parallel or antiparallel to its orbital angular momentum ($j=\ell+s$ vs. $j=\ell-s$). 
Most of the time the monopole component is an average over all possible spin orientations, so the tensor component does not contribute for the filled orbits. However, near the Fermi surfaces where the spin-orbit force splits the orbits, and the $j=\ell +s$ orbit fills first altering the mean field. Indeed residual effective force between the valence nucleons, beyond that represented by the mean field, is very well described by the tensor force \cite{Otsuka:2009qs}. 

Clearly much progress took place in  nuclear structure theory within the last decade.  Recent work by Erler, {\em et al.} \cite{erler}, who recently calculated nuclear driplines using density functional theory with several Skyrme interactions, nicely characterizes this tremendous progress. 

Neutrinoless double beta decay is the most powerful experimental tool to explore the Majorana nature of the neutrinos. Understanding nuclear matrix elements is crucial to be able to fully exploit this tool. Nuclear matrix elements for the neutrinoless double beta decay had been calculated using several different methods, including QRPA \cite{Simkovic:2007vu}, interacting shell model \cite{Caurier:2007wq} and interacting boson model \cite{Barea:2009zza}. The origin of the discrepancies between these calculations is being investigated.  
One of the open problems is deciding on the value of the axial coupling constant, $g_A$. It is known that $g_A$ is strongly renormalized in two-neutrino double beta decays. However, in most of the calculations of the neutrinoless double beta decay the free-nucleon value is used. Since the decay rate is proportional to $(g_A)^4$, this could lead to a sizable effect and needs to be explored further \cite{Barea:2012zz}. 



\subsection{Subbarrier fusion}

Channel-coupling effects in fusion reactions below the Coulomb barrier is well understood \cite{Balantekin:1997yh}, especially for asymmetric systems. Nevertheless there are several outstanding problems. The surface diffuseness parameter of the nuclear potential can be unambiguously determined from the quasi-elastic scattering data \cite{Hagino:2004du}. However, the value of the surface diffuseness parameter needed to fit the subbarrier fusion data is almost twice as much as the value determined from quasi-elastic scattering. Recently, subbarrier fusion cross sections were measured at energies significantly below the Coulomb barrier \cite{Jiang:2002zz,Stefanini:2008zz,nanda}. It was found that standard couple-channel calculations overestimate the fusion cross sections at deep subbarrier energies. This behavior can be explained equally well with two distinct models: either with a repulsive core leading to a shallow potential pocket 
\cite{Misicu:2006du} or by neck formation between colliding nuclei \cite{Ichikawa:2007ei}. Since the physics principles  behind these two models are quite different, the origin of the subbarrier fusion hindrance and its connection to the diffusion parameter choice need to explored further. A recent summary was given in Ref. \cite{Hagino:2012cu}. 

\section{Physics with rare isotopes}

Currently many major accelerator projects around the world, at different stages of construction and operation, aim to explore the physics 
of exotic nuclei. Scientific motivations for pursuing physics with rare isotope beams can be categorized under four groups:
\begin{enumerate}
\item {\bf Nuclear Astrophysics:} Origin of new elements, rare isotopes powering stellar explosions, neutron star crust. 
\item {\bf Nuclear Structure Physics:} Exploring limits of nuclear stability, new shapes, and new collective behavior. 
\item {\bf Fundamental Symmetries:} Use of rare isotopes as laboratories where symmetry violations are amplified. 
\item {\bf Nuclear Applications:} Materials, medical physics, reactors, $\cdots$. 
\end{enumerate}

I will briefly explore a few of the recent developments in some of these areas. A comprehensive and necessarily much longer list  is beyond the scope of this contribution. 

\subsection{Nuclear Astrophysics} 

Understanding how stars shine has been one of the main goals of nuclear astrophysics since the seminal work of Hans Bethe. Recent solar neutrino measurements helped us to understand the structure our Sun better, however many questions remain. CNO neutrinos coming from the Sun are not yet measured. Measurement of these neutrinos would not be just a technical exercise in identifying rare signals: Recently there has been considerable progress in modeling the Sun, especially its metallicity, namely abundances of elements heavier than hydrogen and helium. An improved analysis of the solar abundances gives the ratio of this quantity to hydrogen abundance to be $(Z/X)_{\odot} = 0.0178$   \cite{Asplund:2009fu}, as compared to the previously established value of $(Z/X)_{\odot} = 0.0229$ \cite{Grevesse:1998bj}. Using these rates and recent reevaluation of the nuclear reaction rates in the Sun \cite{Adelberger:2010qa}, new solar model calculations 
\cite{Serenelli:2011py} suggest that the difference between old and new metallicities is hard to distinguish using already measured pp chain neutrino fluxes, however it is quite manifest in the yet-to-be measured CNO neutrino fluxes. From the experimental side of laboratory nuclear astrophysics three reactions stand out for a more precise measurement: The $^3$He($\alpha,\gamma$)$^7$Be reaction contributes to the main nuclear physics uncertainty for the Sun and the Big-Bang nucleosynthesis. The $^{14}$N(p,$\gamma$)$^{15}$O reaction is the dominant  reaction for the CNO burning. Finally the rate of $^{12}$C($\alpha,\gamma$)$^{16}$O reaction is the determining quantity for the carbon-to-oxygen ratio in the Universe, crucial for the proper development of much organic material. 

\subsection{r-process nucleosynthesis}

Nuclei at  all three r-process peaks are observed \cite{Roederer:2012ve}. However, the site of r-process nucleosynthesis is still an open question. The high-temperature, high-entropy region outside the newly-formed neutron star in a core-collapse supernova was suggested to be an r-process site \cite{Meyer:1992zz}. The nuclei with A $\sim$ 90 - 110 produced by the charged-particle reactions in the neutrino-driven winds and the heavier r-process nuclei do not seem to be strongly coupled, suggestion the presence of several possible sites \cite{Qian:2007vq}. 
The neutrino-driven wind blown from the neutron star drives appreciable mass loss from its surface. This wind could be the site where the r-process occurs \cite{Takahashi:1994yz}. Our present understanding of neutrino-driven winds was summarized in Ref. \cite{Arcones:2012wj}. Significant input from nuclear physics is also needed in r-process nucleosynthesis calculations \cite{Petermann:2008cj}. 

Whether the core-collapse supernovae are sites for the r-process nucleosynthesis or not, they are very much neutrino-dominated dynamical systems \cite{Balantekin:2003ip}. 
Recently much progress in understanding such supernovae took place. In particular, development of two- and three-dimensional models revealed a complex interplay between turbulence, neutrino physics and thermonuclear reactions \cite{Bruenn:2012mj,Mueller:2010nf,Brandt:2010xa}.  The sheer number of neutrinos emitted in the cooling of the proto-neutron star ($\sim 10^{58}$), uniquely characterizes core-collapse supernovae as venues where neutrino-neutrino interactions could play a dominant role. Such collective neutrino oscillations were extensively studied \cite{Duan:2010bg,Raffelt:2012kt}. They could significantly impact r-process nucleosynthesis yields 
\cite{Balantekin:2004ug,Duan:2010af}. 

Part of the research program with rare ion beams is to understand the r-process. One of the physics goals is studying beta-decays of nuclei both at and far-from stability:  One needs half-lifes at the r-process ladders (N=50, 82, 126) where abundances peak and accurate values of initial and final state energies. Both of those goals can be achieved with direct measurements using rare ion beams. 
In general understanding the spin-isospin response of a broad range of nuclei to a variety of probes is crucial for not only for the r-process nucleosynthesis, but also for many other astrophysics applications. Much progress was made in measuring matrix elements of the Gamow-Teller operator $\vec{\sigma} \> \vec{\tau}$ between the initial and final states using inverse kinematics \cite{Thies:2012xc}. 

\subsection{Electric dipole moment}

Searching for the time-reversal violating permanent electric dipole moments of nuclei is among the fundamental symmetries tests that 
can be carried out with rare ions. It was noted some time ago that mixing of nearly degenerate opposite-parity ground-state doublets in  deformed nuclei could lead to electric dipole and magnetic quadrupole moments \cite{Haxton:1983dq}. For example reflection asymmetric, octupole deformed long-lived odd-mass isotopes of Rn, Fr, Ra, Ac and Pa may posses enhanced electric dipole moments \cite{Spevak:1996tu}. 
In particular one expects a fairly large Schiff moment for the nucleus $^{225}$Ra \cite{Dobaczewski:2005hz}.  Clearly there are many experimental opportunities for exploiting rare isotopes in such novel symmetry tests. Successful experiments in this realm would significantly broaden the scientific base of the rare isotope facilities establishing a bridge to the atomic and particle physics. 

\section{QCD and nuclear physics}

Quantum chromodynamics (QCD) provides a theoretical framework for understanding how strongly-interacting matter behaves. Even though it is sometimes more convenient to use a phenomenological description of the nuclear properties, one expects that a proper description of the nuclear physics will eventually connect to the QCD. Already, as described in the preceding paragraphs, effective theories which characterize the low-energy behavior had much success in nuclear structure physics. 

Experiments with relativistic heavy ions explore the dynamics of nuclear collisions at energies much higher than those so far addressed here. 
It is easy to experimentally demonstrate that nuclear collisions at high energies are not simply a superposition of nucleon collisions (see e.g. 
Ref. \cite{Adler:2006hu}). One may imagine describing such collisions using Glauber theory. However, Glauber formula and its extensions represent multiple scatterings in the target, but do not take into account the emergent properties of the quark-gluon system for which there are strong experimental hints. A better formulation would enormously help cosmic ray physics.

Experiments at the relativistic heavy ion collider indicate that the quark gluon plasma behaves like a strongly-coupled perfect liquid rather than a weakly-coupled plasma as one would naively expect from the weak-coupling limit of the QCD.  (For a recent summary of the experimental landscape see \cite{Jacak:2012dx}). 
Shear viscosity is the force per unit area created by a shear flow with a transverse gradient. A "good" fluid would have a small viscosity. Using the uncertainty relation, one can argue that there has to exist a lower limit on viscosity \cite{Danielewicz:1984ww}. It is possible to show that, at least in the strong coupling limit of N=4 supersymmetric QCD, the viscosity normalized to the entropy density is $\eta/s = \hbar/(4 \pi k_B)$, where $k_B$ is Boltzman's constant 
\cite{Policastro:2001yc}, equal to the lower limit conjectured from the uncertainty principle arguments. (A more comprehensive discussion is given in Ref. \cite{Schaefer:2009px}). Theoretical analysis of the recent measurements of the viscosity at relativistic heavy-ion collisions 
\cite{Aamodt:2010pa,Agakishiev:2011fs,Adare:2011tg} suggests that the observed value is very close to this lower limit 
\cite{Song:2010mg}. 

\section*{Acknowledgments}

This work was supported in part by the U.S. National Science Foundation Grant No. PHY-1205024 and in part by the University of Wisconsin Research Committee with funds granted by the Wisconsin Alumni Research Foundation.


\begin{thebibliography}{99}

\bibitem{nas1}
http://www.nap.edu/catalog.php?record$\_$id=13438. 

\bibitem{Epelbaum:2011md} 
  E.~Epelbaum, H.~Krebs, D.~Lee and U.~-G.~Meissner,
  Phys.\ Rev.\ Lett.\  {\bf 106}, 192501 (2011)
  [arXiv:1101.2547 [nucl-th]]; 
  E.~Epelbaum, H.~Krebs, T.~Lahde, D.~Lee and U.~-G.~Meissner,
  arXiv:1208.1328 [nucl-th].
  
\bibitem{wiringa}  
R. Wiringa, http://www.int.washington.edu/talks/WorkShops/int$\_$12$\_$3/.

\bibitem{Maris:2011as} 
  P.~Maris, J.~P.~Vary, P.~Navratil, W.~E.~Ormand, H.~Nam and D.~J.~Dean,
  Phys.\ Rev.\ Lett.\  {\bf 106}, 202502 (2011)
  [arXiv:1101.5124 [nucl-th]].

\bibitem{Steiner:2011ft} 
  A.~W.~Steiner and S.~Gandolfi,
  Phys.\ Rev.\ Lett.\  {\bf 108}, 081102 (2012)
  [arXiv:1110.4142 [nucl-th]].
  
\bibitem{Gandolfi:2011xu} 
  S.~Gandolfi, J.~Carlson and S.~Reddy,
  Phys.\ Rev.\ C {\bf 85}, 032801 (2012)
  [arXiv:1101.1921 [nucl-th]].
  
\bibitem{Otsuka:2009cs} 
  T.~Otsuka, T.~Suzuki, J.~D.~Holt, A.~Schwenk and Y.~Akaishi,
  Phys.\ Rev.\ Lett.\  {\bf 105}, 032501 (2010)
  [arXiv:0908.2607 [nucl-th]].
  
\bibitem{Hebeler:2009iv} 
  K.~Hebeler and A.~Schwenk,
  Phys.\ Rev.\ C {\bf 82}, 014314 (2010)
  [arXiv:0911.0483 [nucl-th]].
  
\bibitem{Hebeler:2010xb} 
  K.~Hebeler, S.~K.~Bogner, R.~J.~Furnstahl, A.~Nogga and A.~Schwenk,
  Phys.\ Rev.\ C {\bf 83}, 031301 (2011)
  [arXiv:1012.3381 [nucl-th]].
  
\bibitem{Navratil:2011sa} 
  P.~Navratil, R.~Roth and S.~Quaglioni,
  Phys.\ Lett.\ B {\bf 704}, 379 (2011)
  [arXiv:1105.5977 [nucl-th]].

\bibitem{Suzuki:2003fw} 
  T.~Suzuki, R.~Fujimoto and T.~Otsuka,
  Phys.\ Rev.\ C {\bf 67}, 044302 (2003).
  
\bibitem{Suzuki:2012aa} 
  T.~Suzuki, A.~B.~Balantekin and T.~Kajino,
  Phys.\ Rev.\ C {\bf 86}, 015502 (2012)
  [arXiv:1204.4231 [nucl-th]].
  
\bibitem{Otsuka:2005zz} 
  T.~Otsuka, T.~Suzuki, R.~Fujimoto, H.~Grawe and Y.~Akaishi,
  Phys.\ Rev.\ Lett.\  {\bf 95}, 232502 (2005).
  
\bibitem{Otsuka:2009qs} 
  T.~Otsuka, T.~Suzuki, M.~Honma, Y.~Utsuno, N.~Tsunoda, K.~Tsukiyama and M.~Hjorth-Jensen,
  Phys.\ Rev.\ Lett.\  {\bf 104}, 012501 (2010)
  [arXiv:0910.0132 [nucl-th]].
  
\bibitem{erler}
J. Erler, {\em et al.}, Nature {\bf 486}, 509 (2012).  
  
\bibitem{Simkovic:2007vu} 
  F.~Simkovic, A.~Faessler, V.~Rodin, P.~Vogel and J.~Engel,
  Phys.\ Rev.\ C {\bf 77}, 045503 (2008)
  [arXiv:0710.2055 [nucl-th]].

\bibitem{Caurier:2007wq} 
  E.~Caurier, J.~Menendez, F.~Nowacki and A.~Poves,
  Phys.\ Rev.\ Lett.\  {\bf 100}, 052503 (2008)
  [arXiv:0709.2137 [nucl-th]].
  
\bibitem{Barea:2009zza} 
  J.~Barea and F.~Iachello,
  Phys.\ Rev.\ C {\bf 79}, 044301 (2009).
  
\bibitem{Barea:2012zz} 
  J.~Barea, J.~Kotila and F.~Iachello,
  Phys.\ Rev.\ Lett.\  {\bf 109}, 042501 (2012).
  
\bibitem{Balantekin:1997yh} 
  A.~B.~Balantekin and N.~Takigawa,
  Rev.\ Mod.\ Phys.\  {\bf 70}, 77 (1998)
  [nucl-th/9708036].
  
\bibitem{Hagino:2004du} 
  K.~Hagino and A.~B.~Balantekin,
  Phys.\ Rev.\ A {\bf 70}, 032106 (2004)
  [nucl-th/0404092].
 
\bibitem{Jiang:2002zz} 
  C.~L.~Jiang, H.~Esbensen, K.~E.~Rehm, B.~B.~Back, R.~V.~F.~Janssens, J.~A.~Caggiano, P.~Collon and J.~Greene {\it et al.},
  Phys.\ Rev.\ Lett.\  {\bf 89}, 052701 (2002); 
  C.~L.~Jiang, K.~E.~Rehm, R.~V.~F.~Janssens, H.~Esbensen, I.~Ahmad, B.~B.~Back, P.~Collon and C.~N.~Davids {\it et al.},
  Phys.\ Rev.\ Lett.\  {\bf 93}, 012701 (2004)
  [nucl-ex/0402019];  
  C.~L.~Jiang, B.~B.~Back, H.~Esbensen, R.~V.~F.~Janssens and K.~E.~Rehm,
  Phys.\ Rev.\ C {\bf 73}, 014613 (2006)
  [nucl-ex/0508001]; 
  C.~L.~Jiang, K.~E.~Rehm, H.~Esbensen, B.~B.~Back, R.~V.~F.~Janssens, P.~Collon, C.~M.~Deibel and B.~DiGiovine {\it et al.},
  Phys.\ Rev.\ C {\bf 81}, 024611 (2010).
  
\bibitem{Stefanini:2008zz} 
  A.~M.~Stefanini, G.~Montagnoli, R.~Silvestri, S.~Beghini, L.~Corradi, S.~Courtin, E.~Fioretto and B.~Guiot {\it et al.},
  Phys.\ Rev.\ C {\bf 78}, 044607 (2008).
  
\bibitem{nanda}  
M. Dasgupta, D. J. Hinde, A. Diaz-Torres, B. Bouriquet, Catherine I. Low, G. J. Milburn, and J. O. Newton, Phys. Rev. Lett. {\bf 99}, 192701 (2007).    
  
\bibitem{Misicu:2006du} 
  S.~Misicu and H.~Esbensen,
  Phys.\ Rev.\ Lett.\  {\bf 96}, 112701 (2006)
  [nucl-th/0602064].

\bibitem{Ichikawa:2007ei} 
  T.~Ichikawa, K.~Hagino and A.~Iwamoto,
  Phys.\ Rev.\ C {\bf 75}, 064612 (2007)
  [arXiv:0704.2827 [nucl-th]].

\bibitem{Hagino:2012cu} 
  K.~Hagino and N.~Takigawa, 
  Prog. Theor. Phys. {\bf 128}, 1061 (2012)  
  [arXiv:1209.6435 [nucl-th]].
  
\bibitem{Asplund:2009fu} 
  M.~Asplund, N.~Grevesse, A.~J.~Sauval and P.~Scott,
  Ann.\ Rev.\ Astron.\ Astrophys.\  {\bf 47}, 481 (2009)
  [arXiv:0909.0948 [astro-ph.SR]].
  
\bibitem{Grevesse:1998bj} 
  N.~Grevesse and A.~J.~Sauval,
  Space Sci.\ Rev.\  {\bf 85}, 161 (1998).
  
\bibitem{Adelberger:2010qa} 
  E.~G.~Adelberger, A.~B.~Balantekin, D.~Bemmerer, C.~A.~Bertulani, J.~-W.~Chen, H.~Costantini, M.~Couder and R.~Cyburt {\it et al.},
  Rev.\ Mod.\ Phys.\  {\bf 83}, 195 (2011)
  [arXiv:1004.2318 [nucl-ex]].
  
\bibitem{Serenelli:2011py} 
  A.~M.~Serenelli, W.~C.~Haxton and C.~Pena-Garay,
  Astrophys.\ J.\  {\bf 743}, 24 (2011)
  [arXiv:1104.1639 [astro-ph.SR]].
  
\bibitem{Roederer:2012ve} 
  I.~U.~Roederer and J.~E.~Lawler,
  Astrophys.\ J.\  {\bf 750}, 76 (2012)
  [arXiv:1204.3901 [astro-ph.SR]].
  
\bibitem{Meyer:1992zz} 
  B.~S.~Meyer, G.~J.~Mathews, W.~M.~Howard, S.~E.~Woosley and R.~D.~Hoffman,
  Astrophys.\ J.\  {\bf 399}, 656 (1992).
  
\bibitem{Qian:2007vq} 
  Y.~-Z.~Qian and G.~J.~Wasserburg,
  Phys.\ Rept.\  {\bf 442}, 237 (2007)
  [arXiv:0708.1767 [astro-ph]].
  
\bibitem{Takahashi:1994yz} 
  K.~Takahashi, J.~Witti and H.~-T.~Janka,
  Astron.\ Astrophys.\  {\bf 286}, 857 (1994); 
  Y.~Z.~Qian and S.~E.~Woosley,
  Astrophys.\ J.\  {\bf 471}, 331 (1996)
  [astro-ph/9611094].
    
\bibitem{Arcones:2012wj} 
  A.~Arcones and F.~-K.~Thielemann,
  J.\ Phys.\ G {\bf 40}, 013201 (2013)
  [arXiv:1207.2527 [astro-ph.SR]].
  
\bibitem{Petermann:2008cj} 
  I.~Petermann, K.~Langanke, G.~Martinez-Pinedo, P.~v.~Neumann-Cosel, F.~Nowacki and A.~Richter,
  Phys.\ Rev.\ C {\bf 81}, 014308 (2010)
  [arXiv:0812.0968 [astro-ph]].
  
\bibitem{Balantekin:2003ip} 
  A.~B.~Balantekin and G.~M.~Fuller,
  J.\ Phys.\ G {\bf 29}, 2513 (2003)
  [astro-ph/0309519].
  
\bibitem{Bruenn:2012mj} 
  S.~W.~Bruenn, A.~Mezzacappa, W.~R.~Hix, E.~J.~Lentz, O.~E.~B.~Messer, E.~J.~Lingerfelt, J.~M.~Blondin and E.~Endeve {\it et al.},
  arXiv:1212.1747 [astro-ph.SR];  
  E.~J.~Lentz, A.~Mezzacappa, O.~E.~Bronson Messer, W.~R.~Hix and S.~W.~Bruenn,
  Astrophys.\ J.\  {\bf 760}, 94 (2012)
  [arXiv:1206.1086 [astro-ph.SR]].
  
\bibitem{Mueller:2010nf} 
  B.~Muller, H.~-T.~Janka and H.~Dimmelmeier,
  Astrophys.\ J.\ Suppl.\  {\bf 189}, 104 (2010)
  [arXiv:1001.4841 [astro-ph.SR]]; 
  B.~Muller, H.~-T.~Janka and A.~Marek,
  Astrophys.\ J.\  {\bf 756}, 84 (2012)
  [arXiv:1202.0815 [astro-ph.SR]].
  
\bibitem{Brandt:2010xa} 
  T.~D.~Brandt, A.~Burrows, C.~D.~Ott and E.~Livne,
  Astrophys.\ J.\  {\bf 728}, 8 (2011)
  [arXiv:1009.4654 [astro-ph.HE]]; 
  J.~Nordhaus, A.~Burrows, A.~Almgren and J.~Bell,
  Astrophys.\ J.\  {\bf 720}, 694 (2010)
  [arXiv:1006.3792 [astro-ph.SR]].
  
\bibitem{Duan:2010bg} 
  H.~Duan, G.~M.~Fuller and Y.~-Z.~Qian,
  Ann.\ Rev.\ Nucl.\ Part.\ Sci.\  {\bf 60}, 569 (2010)
  [arXiv:1001.2799 [hep-ph]].
  
\bibitem{Raffelt:2012kt} 
  G.~Raffelt,
  arXiv:1201.1637 [astro-ph.SR].
  
\bibitem{Balantekin:2004ug} 
  A.~B.~Balantekin and H.~Yuksel,
  New J.\ Phys.\  {\bf 7}, 51 (2005)
  [astro-ph/0411159].
  
\bibitem{Duan:2010af} 
  H.~Duan, A.~Friedland, G.~C.~McLaughlin and R.~Surman,
  J.\ Phys.\ G {\bf 38}, 035201 (2011)
  [arXiv:1012.0532 [astro-ph.SR]]; 
  S.~Chakraborty, S.~Choubey, S.~Goswami and K.~Kar,
  JCAP {\bf 1006}, 007 (2010)
  [arXiv:0911.1218 [hep-ph]].

\bibitem{Thies:2012xc} 
  J.~H.~Thies, P.~Puppe, T.~Adachi, M.~Dozono, H.~Ejiri, D.~Frekers, H.~Fujita and Y.~Fujita {\it et al.},
  Phys.\ Rev.\ C {\bf 86}, 054323 (2012);  
  J.~H.~Thies, T.~Adachi, M.~Dozono, H.~Ejiri, D.~Frekers, H.~Fujita, Y.~Fujita and M.~Fujiwara {\it et al.},
  Phys.\ Rev.\ C {\bf 86}, 044309 (2012).
  
\bibitem{Haxton:1983dq} 
  W.~C.~Haxton and E.~M.~Henley,
  Phys.\ Rev.\ Lett.\  {\bf 51}, 1937 (1983).
  
\bibitem{Spevak:1996tu} 
  V.~Spevak, N.~Auerbach and V.~V.~Flambaum,
  Phys.\ Rev.\ C {\bf 56}, 1357 (1997)
  [nucl-th/9612044].
  
\bibitem{Dobaczewski:2005hz} 
  J.~Dobaczewski and J.~Engel,
  Phys.\ Rev.\ Lett.\  {\bf 94}, 232502 (2005)
  [nucl-th/0503057].
  
\bibitem{Adler:2006hu} 
  S.~S.~Adler {\it et al.}  [PHENIX Collaboration],
  Phys.\ Rev.\ Lett.\  {\bf 96}, 202301 (2006)
  [nucl-ex/0601037].
  
\bibitem{Jacak:2012dx} 
  B.~V.~Jacak and B.~Muller,
  Science {\bf 337}, 310 (2012).
 
\bibitem{Danielewicz:1984ww} 
  P.~Danielewicz and M.~Gyulassy,
  Phys.\ Rev.\ D {\bf 31}, 53 (1985).

\bibitem{Policastro:2001yc} 
  G.~Policastro, D.~T.~Son and A.~O.~Starinets,
  Phys.\ Rev.\ Lett.\  {\bf 87}, 081601 (2001)
  [hep-th/0104066].

\bibitem{Schaefer:2009px} 
  T.~Schafer and C.~Chafin,
  Lect.\ Notes Phys.\  {\bf 836}, 375 (2012)
  [arXiv:0912.4236 [cond-mat.quant-gas]]; 
  T.~Schafer and D.~Teaney,
  Rept.\ Prog.\ Phys.\  {\bf 72}, 126001 (2009)
  [arXiv:0904.3107 [hep-ph]].
 
\bibitem{Aamodt:2010pa} 
  KAamodt {\it et al.}  [ALICE Collaboration],
  Phys.\ Rev.\ Lett.\  {\bf 105}, 252302 (2010)
  [arXiv:1011.3914 [nucl-ex]].
  
\bibitem{Agakishiev:2011fs} 
  H.~Agakishiev {\it et al.}  [STAR Collaboration],
  Phys.\ Lett.\ B {\bf 704}, 467 (2011)
  [arXiv:1106.4334 [nucl-ex]].
  
\bibitem{Adare:2011tg} 
  A.~Adare {\it et al.}  [PHENIX Collaboration],
  Phys.\ Rev.\ Lett.\  {\bf 107}, 252301 (2011)
  [arXiv:1105.3928 [nucl-ex]].
  
\bibitem{Song:2010mg} 
  H.~Song, S.~A.~Bass, U.~Heinz, T.~Hirano and C.~Shen,
  Phys.\ Rev.\ Lett.\  {\bf 106}, 192301 (2011)
  [Erratum-ibid.\  {\bf 109}, 139904 (2012)]
  [arXiv:1011.2783 [nucl-th]]; 
  C.~Shen, S.~A.~Bass, T.~Hirano, P.~Huovinen, Z.~Qiu, H.~Song and U.~Heinz,
  J.\ Phys.\ G {\bf 38}, 124045 (2011)
  [arXiv:1106.6350 [nucl-th]].
  
\end{thebibliography}
\end{document}